# Pharmacokinetics and Molecular Docking studies of Plant-Derived Natural Compounds to Exploring Potential Anti-Alzheimer Activity

Aftab Alam[†1], Naaila Tamkeen[†1], Nikhat Imam[2], Anam Farooqui[1], Mohd Murshad Ahmed[1], Shahnawaz Ali[1], Md Zubbair Malik[1] and Romana Ishrat[1]*

[1]Centre for Interdisciplinary Research in Basic Science, Jamia Millia Islamia, New Delhi, India- 110025
[2]Institute of Computer Science and Information Technology, Magadh University, Bodhgaya.

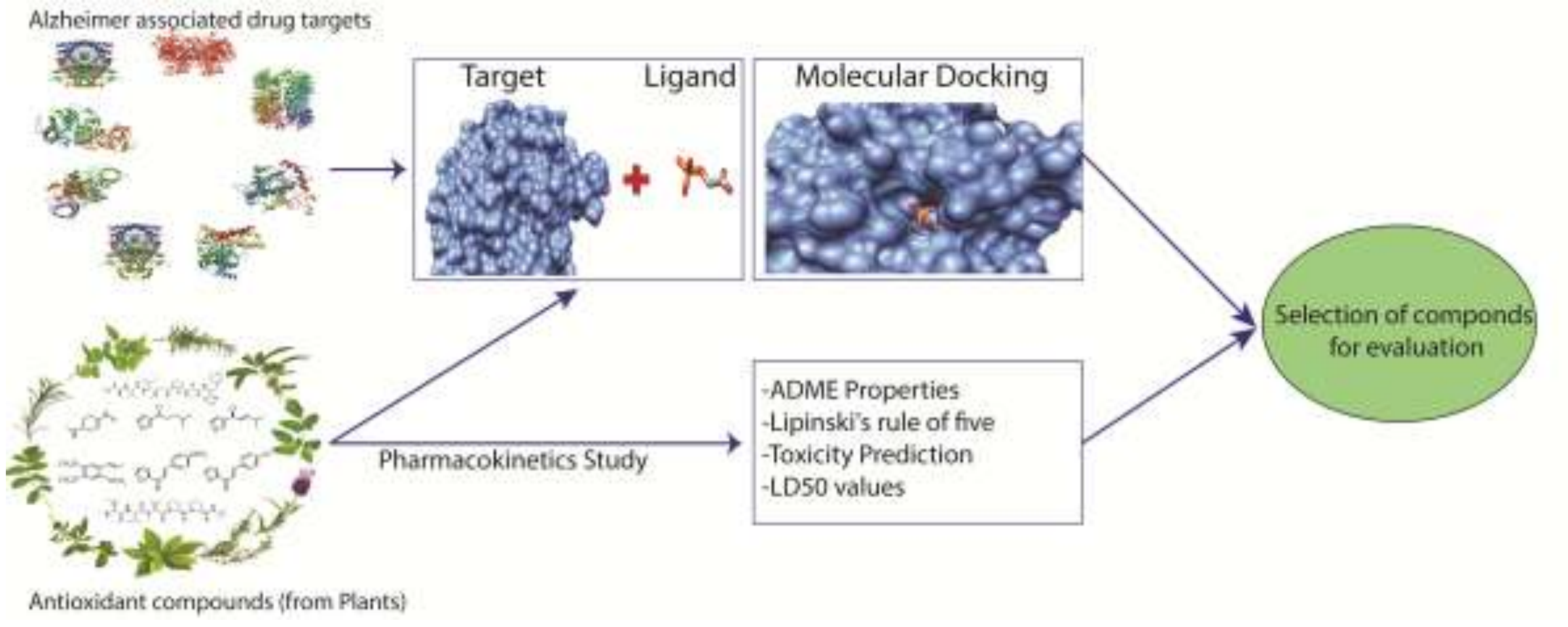

†= AA and NT contributed equally to this work.

***Corresponding Author:**

**Dr. Romana Ishrat**

Assistant Professor
Computational & Bioinformatics lab
Centre for Interdisciplinary Research in Basic Science
Jamia Millia Islamia
New Delhi-110025, India
E-mail: rishrat@jmi.ac.in

# Abstract

Alzheimer's disease (AD) is the leading cause of dementia, accounts for 60 to 80 percent cases**.** Two main factors called β-amyloid (Aβ) plaques and tangles are prime suspects in damaging and killing nerve cells. However, oxidative stress, the process which produces free radicals in cells, is believed to promote its progression to the extent that it may responsible for the cognitive and functional decline observed in AD. As of today there are few FDA approved drugs in the market for treatment, but their cholinergic adverse effect, potentially distressing toxicity and limited targets in AD pathology limits their use. Therefore, it is crucial to find an effective compounds to combat AD. We choose 45 plant-derived natural compounds that have antioxidant properties to slow down disease progression by quenching free redicals or promoting endogenous antioxidant capacity. However, we performed molecular docking studies to investigate the binding interactions between natural compounds and 13 various anti-Alzheimer drug targets. Three known Cholinesterase inhibitors (Donepezil, Galantamine and Rivastigmine) were taken as reference drugs over natural compounds for comparison and drug-likeness studies. Few of these compounds showed good inhibitory activity besides anti-oxidant activity. Most of these compounds followed pharmacokinetics properties that make them potentially promising drug candidates for the treatment of Alzheimer’s disease.

Graphical Abstract : Pharmacokinatics and Molecular docking studies of 45 natural antioxidant compounds with most known Alzheimer asscociated targets.

# Introduction

Alzheimer's disease  is one of the most common neurodegenerative disorders that normally cause dementia and affect the middle to old-aged persons, around one in four individuals over the age of 85[1]. Alzheimer's is a progressive disease, where dementia steadily gets worse over the years. The World Alzheimer Report 2015 led by King's College London found that there are currently around 46.8 million people living with dementia around the world, with numbers projected to nearly double every 20 years, increasing to 74.7 million by 2030 and 131.5 million by 2050 (Fig. 1). At the country level, ten countries are home to over a million people with dementia in 2015: China (9.5 million), US (4.2 million), India (4.1 million), Japan (3.1 million), Brazil (1.6 million), Germany (1.6 million), Russia (1.3 million), Italy (1.2 million), Indonesia (1.2 million) and France (1.2 million). Thus, this condition will bring gigantic financial and personal burdens to current and future generations. In order to deal with this problem, effective therapeutic and preventive interventions should be developed urgently.

Figure 1 People living with dementia around the World: 9.4 million people of Americas, 4.0 million of African region, 10.5 *million* of Europian countries and 22.9 *million* people of Asia are living with dementia in 2015. This no. will reach 15.8 and 29.9 *million* in Americas,7.0 and 15.8 million in Africa, 13.4 and 18.6 *million* in Europe and 38.5 and 67.2 *million* in Asia in 2030 and 2050 respectively. (*World Alzheimer Report* )

There are no such drugs/treatments available that can cure AD or any other common type of dementia completely. However, medications have been developed for Alzheimer's disease that can temporarily attenuate the symptoms, or delay it progression. The U.S. Food and Drug Administration (FDA) have approved two medications-cholinesterase inhibitors and Memantine. Over the past decade, much of the research on Alzheimer disease (AD) has focused on oxidative stress mechanisms and its importance in disease pathogenesis. The net effect of oxygen radicals is damaging, such damage present in AD includes advanced glycation end products [2], nitration [3], lipid peroxidation adduction products [4-5] as well as carbonyl-modified neurofilament protein and free carbonyls [6-7]. Significantly, this damage involves all neurons at risk to death in AD, not just those containing neurofibrillary tangles.

Nature has gifted us lots of natural remedies in the form of  fruits, leaves, bark, vegetables and nuts, etc. The various ranges of  bioactive nutrients present in these natural products play a vital role in prevention and cure of various neurodegenerative diseases, such as AD, Parkinson's disease and other neuronal dysfunctions. Previous studies  suggested that the  naturally occurring  phytochemicals, such as polyphenolic antioxidants found in fruits, vegetables, herbs and nuts, may potentially hinder neurodegeneration, and improve memory and cognitive functions.

In our study, we chose 45 natural compounds (antioxident properties) of various plants from different databases (Table 1) for molecular docking and pharmacokinetics studies. Recent studies have demonstrated that the natural compounds and their derivatives possess wide range of biological activities like anti- tubercular, antifungal, antibacterial anti-malarial, anti-inflammatory and antioxidant activities [8-10].

Table 1: List of studied compounds and soarces

| S. no. | Compound | Molecular Formula | Plants and food sources |
|---|---|---|---|
| 1. | Vitamin C | $C_6H_8O_6$ | Kakadu plum, Camu Camu, Acerola, citrus fruits (such as oranges, sweet lime, etc.), green peppers, broccoli, green leafy vegetables, black currants, strawberries, blueberries, seabuckthorn, raw cabbage and tomatoes. |
| 2. | Vitamin E | $C_{29}H_{50}O_2$ | Wheat germ, seabuckthorn, nuts, seeds, whole grains, green leafy vegetables, kiwifruit, vegetable oil and fish-liver oil |
| 3. | Coenzyme Q10 | $C_{59}H_{90}O_4$ | Sardine, Mackrel ,beef, pork, chicken heart, chicken liver, rapeseed oil, soybean oil, sesame oil, peanuts, parsley, perilla, broccoli, grapes, cauliflower, avocados etc |
| 4. | Iodide | $I^-$ | Sea vegetables, cranberries, yogurt, navy beans, strawberries, raw cheese, potatoes etc. |
| 5. | Melatonin | $C_{13}H_{16}N_2O_2$ | Cherries, bananas, grapes, rice, cereals, herbs, plums, olive oil, wine, beer, pineapple, oranges etc. |
| 6. | Alpha carotene | $C_{40}H_{56}$ | Carrots, sweet potatoes, pumpkin, winter squash, broccoli, green beans, green peas, spinach, turnip greens, collards, leaf lettuce, avocado, parsley etc. |
| 7. | Astaxanthin | $C_{40}H_{52}O_4$ | Microalgae, yeast, Salmon, Trout, Krill, Shrimp, Crayfish, crustaceans etc. |
| 8. | Beta carotene | $C_{40}H_{56}$ | Butternut squash, carrots, orange bell peppers, pumpkins, kale, peaches, apricots, mango, turnip greens, broccoli, spinach, and sweet potatoes. |
| 9. | Canthaxanthin | $C_{40}H_{52}O_2$ | Pacific salmon, green algae, crustacean and fishes like Crap, Golden mullet, Seabream and Trush wasse. |
| 10. | Lutein | $C_{40}H_{56}O_2$ | Spinach, kale, Swiss chard, collard greens, beet and mustard greens, endive, red pepper and okra. |
| 11. | Lycopene | $C_{40}H_{56}$ | Cooked red tomato products like canned tomatoes, tomato sauce, tomato juice and garden cocktails, guava, red carrots, watermelons, gac, papayas etc. |
| 12. | Zeaxanthin | $C_{40}H_{56}O_2$ | Paprika, corn, saffron, wolfberries, Spirulina, dark green leafy vegetables, such as kale, spinach, turnip greens, collard greens, romaine lettuce, watercress, Swiss chard and mustard greens. |
| 13. | Apigenin | $C_{15}H_{10}O_5$ | Parsley, celery, celeriac, chamomile tea, flowers of chamomile plants, Navy bean etc |
| 14. | Luteolin | $C_{15}H_{10}O_6$ | *Salvia tomentosa*, celery, broccoli, green pepper, parsley, thyme, dandelion, perilla, chamomile tea, carrots, olive oil, peppermint, rosemary, navel oranges, oregano, seeds of the palm *Aiphanes aculeate* etc. |
| 15. | Tangeritin | $C_{20}H_{20}O_7$ | Tangerine and other citrus peels. |
| 16. | Isorhamnetin | $C_{16}H_{12}O_7$ | Pungent yellow or red onions, Mexican terragon *(Tagetes lucida)* etc. |
| 17. | Kaempferol | $C_{15}H_{10}O_6$ | Apples, grapes, tomatoes, green tea, potatoes, onions, broccoli, Brussels sprouts, squash, cucumbers, lettuce, green beans, peaches, blackberries, raspberries, and spinach. Plants that are known to contain kaempferol include *Aloe vera*, *Coccinia grandis*, *Cuscuta chinensis*, *Euphorbia pekinensis*, *Glycine max*, *Hypericum perforatum*, *Moringa oleifera*, *Rosmarinus officinalis*, *Sambucus nigra*, and *Toona sinensis*. |
| 18. | Myricetin | $C_{15}H_{10}O_8$ | Vegetables, fruits, nuts, berries, tea, and is also found in red wine. |
| 19. | Quercetin | $C_{15}H_{10}O_7$ | Capers, lovage, dock like sorrel, radish leaves, carob fiber, dill, cilantro, Hungarian wax peppers, funnel leaves, onion (red), radicchio, watercress, buckwheat, kale etc. |
| 20. | Eriodictyol | $C_{15}H_{12}O_6$ | In twigs of *Millettia duchesnei*, in *Eupatorium arnottianum*, and its glycosides (eriocitrin) in lemons and rose hips (*Rosa canina*). |
| 21. | Hesperetin | $C_{16}H_{14}O_6$ | Lemons and sweet oranges. |
| 22. | Naringenin | $C_{15}H_{12}O_5$ | Grapefruit, oranges, tomatoes (skin) and in water mint. |
| 23. | Catechin | $C_{15}H_{14}O_6$ | *Uncaria rhynchophylla*, pome fruits, cocoa, prune juice, broad bean pod, acai oil, argan oil, peaches, green tea, vinegar, barley grain, etc |
| 24. | Gallocatechin | $C_{15}H_{14}O_7$ | Green tea, bananas, persimmons and pomegranates. |
| 25. | Epicatechin | | *Uncaria rhynchophylla*, cacao beans, green tea etc. |
| 26. | Epigallocatechin | | St John's wort etc. |
| 27. | Theaflavin | $C_{29}H_{24}O_{12}$ | Black tea etc. |
| 28. | Daidzein | $C_{15}H_{10}O_4$ | Kwao Krua (*Pueraria mirifica*), Kudzu (*Pueraria lobata*), *Maackia amurensis* cell cultures, soybeans, lupin, fava beans, psoralea, *Flemingia vestita*, *F. marophylla* and coffee. |
| 29. | Ganistein | $C_{15}H_{10}O_5$ | Lupin, fava beans, soybeans, kudzu, *Flemingia vestita*, *F. macrophylla*, coffee, *Maackia amurensis* cell cultures etc. |
| 30. | Glycitein | $C_{16}H_{12}O_5$ | Soy food products. |
| 31. | Resveratrol | $C_{14}H_{12}O_3$ | Skin of grapes, blueberries, raspberries, mulberries, lingonberry, senna and concentrated in red wine. |
| 32. | Pterostilbene | $C_{16}H_{16}O_3$ | Almonds, various *Vaccinium* berries, grape leaves, vines and blueberries. |
| 33. | Cyanidin | $C_{15}H_{11}O_6^+$ | Grapes, bilberry, blackberry, blueberry, cherry, cranberry, elderberry, hawthorn, loganberry, açai berry, raspberry, apples, plums, red cabbage and red onion. |
| 34. | Delphinidin | $C_{15}H_{11}O_7^+$ | Cranberries, Concord grapes, pomegranates, bilberries etc |
| 35. | Malvidin | $C_{17}H_{15}O_7$+ | *Primula* plants, blue pimpernel (*Anagallis monelli*), red wine, *Vitis vinifera*, chokeberries (*Aronia sp*), saskatoon berries (*Amelanchier alnifolia*) etc. |
| 36. | Pelargonidin | $C_{15}H_{11}O_5^+$ | Red geraniums (Geraniaceae), *Philodendron* (Araceae), flowers of blue pimpernel (*Anagallis monelli*, Myrsinaceae), raspberries, strawberries, blueberries, blackberries, cranberries, saskatoon berries, chokeberies, plums pomegranates, kidney beans etc. |
| 37. | Peonidin | $C_{16}H_{13}O_6^+$ | Raw cranberries, blueberries, plums, grapes, cherries, purple fleshed sweet potatoes, raw black rice and black bananas. |
| 38. | Petunidin | $C_{16}H_{13}O_7^+$ ($Cl^-$) | Chokeberries (*Aronia sp*), Saskatoon berries (*Amelanchier alnifolia*) or different species of grape (for instance *Vitis vinifera*, or muscadine, *Vitis rotundifolia*), Indigo Rose tomatoes. |
| 39. | Cichoric acid | $C_{22}H_{18}O_{12}$ | *Cichorium intybus* (chicory), *E. purpurea*, dandelion leaves, basil, lemon balm, and aquatic plants, including algae and seagrasses. |
| 40. | Chlorogenic acid | $C_{16}H_{18}O_9$ | Leaves of *Hibiscus sabdariffa*, potatoes, flesh of eggplants,peach, prunes, green coffee bean extract and green tea. |
| 41. | Cinnamic acid | $C_9H_8O_2$ | Oil of cinnamon, balsams such as storax, shea butter, seeds of plants such as brown rice, whole wheat, oats, coffee, apple, artichoke, peanut, orange and pineapple. |
| 42. | Ellagic acid | $C_{14}H_6O_8$ | North American white oak (*Quercus alba*, European red oak (*Quercus robur*). *Myriophyllum spicatum*, medicinal mushroom *Phellinus linteus*, walnuts, pecans, cranberries, raspberries, strawberries, and grapes, as well as distilled beverages, peach and other plant foods. |
| 43. | Gallic acid | $C_7H_6O_5$ | Parasitic plant *Cynomorium coccineum*, the aquatic plant *Myriophyllum spicatum*, the blue-green alga *Microcystis aeruginosa*. oak species, *Caesalpinia mimosoides*, and in the stem bark of *Boswellia dalzielii*, fruits (including strawberries, grapes, bananas), as well as teas, cloves and vinegars. |
| 44. | Rosmarinic acid | $C_{18}H_{16}O_8$ | *Ocimum basilicum* (basil), *Ocimum tenuiflorum* (holy basil), *Melissa officinalis* (lemon balm), *Rosmarinus officinalis* (rosemary), *Origanum majorana* (marjoram), *Salvia officinalis* (sage), thyme and peppermint, (*Prunella vulgaris*) *Heliotropium foertherianum* *Maranta* (*Maranta leuconeura*, *Maranta depressa*) and *Thalia* (*Thalia geniculata*). |
| 45. | Salicyclic acid | $C_7H_6O_3$ | Blackberries, blueberries, cantaloupes, dates, grapes, kiwi fruits, guavas, apricots, green pepper, olives, tomatoes, radish, chicory and mushrooms. Some herbs and spices contain high amounts, Of the legumes, seeds, nuts and cereals, only almonds, water chestnuts and peanuts have significant amounts. |

In our study, all the 45 compounds have antioxidant properties and it has been shown that treatment with these compounds certainly contribute to their neuroprotective effects and it is a potential approach for slowing disease progression. Therefore, we further screened our compounds against Alzheimer, which is caused by oxidative stress and it is one of the main-factors in progression of Alzheimer [11-12]. Natural compounds that have antioxidant properties exhibit their antioxidant effect by quenching free redicals or promoting endogenous antioxidant capacity and some of these stimulate the synthesis of endogenous antioxidant molecules in cell via activation of Nrf/ARE pathways. Therefore, these compounds can be good candidates for the assessment of the AD by scavenging free radicals. Oxidative stress and its sources in AD are illustrated in (Fig. 2).

Figure 2 Antioxidant strategies and sources of oxidative stress in AD.

In our study we also took three cholinesterase inhibitors (Donepezil, Galantamine and Rivastigmine), which are commonly prescribed drugs in Alzheimer, as reference compounds over our natural compounds for drug-likeness studies. Mechanism of action of these drugs by preventing an enzyme called acetylcholinesterase, which breaks down acetylcholine in the brain and enhances antioxidants effects and attenuates oxidative stresses[13-14]. As a result of our investigation, we found that our natural compounds showed promising inhibitory activity also. Some of them were found to have even better activity than prescribed drugs against AD targets. Any compound cannot be directly considered as a drug molecule unless it is validated by several parameters like pharmacokinetic properties, ADME properties, and potential toxicity. Therefore, with the help of various bioinformatics tool, we validated all our compounds.

Molecular docking studies are used to find out the interaction between a ligand/drug and a protein at the atomic level which allows us to characterize the behaviour of our compounds in the binding site of targets as well as to explain fundamental biochemical processes [15]. Each of the natural compounds were docked with all 13 AD associated proteins individually, to determine the best binding affinity using Autodock4.2[16]. Further, these compounds could be useful for the identification and development of new preventive and therapeutic drug against Alzheimer disease .

## Materials and Method

### Natural Compounds Selection

Fourty-five natural compounds were selected as common reported antioxidant natural compounds from various database and literature . All chemical structures of these compunds were sketched in ChemBioDraw Ultra 12.0 (CambrigdeSoft) as shown in (Fig. 3).

Figure 3: Chemical structures of the studied compounds.

### Basic Pharmacokinetics Parameters Calculation

A compound has to be passed through multiple filters to be considered a novel drug. Most of the compounds that fail in pre-clinical trials do so because they do not show the required pharmacological properties to be a drug molecule [17]. Pharmacokinetics properties such as absorption, distribution, metabolism, excretion, and toxicity (ADMET) has play a very crucial role in development of drug design to the final clinical success of a drug candidate [18]. Therefore, prediction of ADMET properties was done earlier with the aim of decreasing the failure rate of the compound for further process in future. Pharmacokinetics properties of natural compounds such as MW (molecular weight), LogP, HBD(number of hydrogen bond donors), HBA (number of hydrogen bond acceptors) , TPSA (topological polar surface area), nrtB (number of rotatable bonds) , nViolation (violations of Lipinski's rule of five) [19] were calculated by DruLito (Drug LiknessTool) (www.niper.gov.in/pi_dev_tools/DruLiToWeb/DruLiTo_index.html). and Molinspiration Online tool (http://www.molinspiration.com/).

### Compound Toxicity Prediction

The compounds toxicities prediction is an important sect of the drug design development process. In silico toxicity assessments are not only faster, but can also reduce the amount of animal experiments. So, calculated $LD_{50}$ values of all our natural compounds. $LD_{50}$ value is the amount of doses given to kill 50% of a test population (lab rats or other animals). It is an index determination of medicine and poison's virulence. **The lower the $LD_{50}$ dose, the greater is the toxicity of the substance**. These $LD_{50}$ values were calculated by an online tool ProTox[20]. We also determined its carcinogenic, mutagenic and skin irritation properties by using Discovery Studio 2.5 (Accelrys Software Inc., San Diego, CA, USA).

## Molecular Docking

### Target preparation

All the 13 Alzheimer disease associated targets were downloaded from Protein Data Bank (PDB) (http://www.rcsb.org/pdb/home/home.do) and are listed in (Table 2). For crystal structure of each target, the crystallographic water molecules were removed, the missing hydrogen atoms were added and the energy level of all the 13 targets was minimised using swiss_PDB viewer tool [21].

Table 2: 13 Alzheimer disease associated targets

| S.No | PDB ID | Protein Name | References |
|---|---|---|---|
| 1 | 1EQG | Cyclooxygenase-1 (COX-1) | [26] |
| 2 | 1MX1 | Human carboxylesterase (hCE-1) | [27] |
| 3 | 1PBQ | N-methyl-D-aspartate (NMDA) | [28] |
| 4 | 1Q5K | Glycogen-synthase-kinase-3β (GSK-3β) | [29] |
| 5 | 1QWC | Nitric oxide synthase (NOS) | [30] |
| 6 | 1UDT | Phosphodiesterase-5 (PD-5) | [31] |
| 7 | 2FV5 | TNF-α converting enzyme (TACE) | [32] |
| 8 | 3BKL | Angiotensin converting enzyme (ACE) | [33] |
| 9 | 3G9N | c-Jun N-terminal kinase (JNK) | [34] |
| 10 | 3QMO | Cyclooxygenase-2 (COX-2) | [35] |
| 11 | 4B0P | Butyrylcholinesterase (BuChE) | [36] |
| 12 | 4DJU | β-Site amyloid precursor protein | [37] |
| 13 | 4EY5 | Acetylcholinesterase (AChE) | [38] |

### Ligand preparation

The structure of all the 45 compounds were drawn in **ChemDraw12** (PerkinElmer Informatics, Waltham, MA, USA) and converted to their 3D form and also geometry of the compounds was optimized in **ChemBio3DUltra12** (PerkinElmer Informatics, Waltham, MA, USA). Finally all compounds were saved in PDB format for further docking studies.

**Target Lignad docking**

Docking studies yielded crucial information concerning the orientation of the inhibitors in the binding pocket of the target proteins. During the molecular docking process, all the natural compounds bound in the groove of their respective targets. Each of the compounds was docked with all the 13 AD associated targets, hence total 585 dockings were performed (Table 3). Then we used 3D sorting method to filter out the best possible compounds from the pool of 45 natural compounds.

Table 3: Binding energies of all the 45 natural compounds and 03 reference drugs docked with 13Alzheimer disease associated targets.

| | 1EQG | 1MX1 | 1PBQ | 1Q5K | 1QWC | 1UDT | 2FV5 | 3BKL | 3G9N | 3QMO | 4B0P | 4DJU | 4EY5 |
|---|---|---|---|---|---|---|---|---|---|---|---|---|---|
| | Affinity (kcal.mol) | Affinity (kcal/mol) | Affinity (kcal.mol) | Affinity (kcal/mol) | Affinity (kcal/mol) | Affinity (kcal/mol) | Affinity (kcal/mol) | Affinity (kcal/mol) | Affinity (kcal/mol) | Affinity (kcal/mol) | Affinity (kcal/mol) | Affinity (kcal/mol) | Affinity (kcal/mol) |
| Alpha carotene | -7.6 | -7.3 | -7.2 | -8.6 | -7.3 | -7.5 | -6.8 | -7.6 | -7.5 | -7.4 | -8 | -8.8 | -7.1 |
| Apigenin | -9.1 | **-9.3** | -8.4 | -8.7 | -9.3 | -9.4 | -8.0 | -8.3 | -7.8 | -7.4 | -7.9 | -8.4 | -8.9 |
| **Astaxanthin** | -8.2 | -7.9 | -7.4 | -8.7 | -6.2 | -7.0 | -8.0 | **-10.0** | -8.1 | -7.4 | -8.7 | **-9.4** | **-9.2** |
| Beta carotene | -7.9 | -7.1 | **-8.7** | -7.4 | -9.0 | -7.7 | -7.6 | -7.8 | -7.3 | -7.9 | -8.5 | -9.3 | -7.1 |
| Canthaxanthin | -8.4 | -7.9 | -7.8 | -8.1 | -9.2 | -7.2 | -8.2 | -7.6 | **-8.7** | -7.5 | -7.9 | -7.4 | -6.9 |
| **Catechin** | -8.7 | **-9.4** | **-8.5** | **-9.7** | **-9.9** | -7.7 | **-10.1** | **-10.4** | **-9.0** | -7.6 | **-10.6** | **-9.5** | **-9.6** |
| Chicoric acid | -8 | -6.6 | -7.3 | -8.5 | -8.6 | -7.1 | -8.5 | -8.6 | -7.4 | -7.2 | -9.6 | -7.3 | -8.2 |
| chlorogenic acid | -7.6 | -7.4 | -6.9 | -8.5 | -4.8 | -6.5 | **-9.8** | -7.8 | -7.5 | -7.5 | -8.3 | -7.4 | -7.6 |
| Cinnamic acid | -5 | -5.5 | -4.2 | -6.2 | -4.7 | -6.5 | -4.7 | -6.1 | -4.7 | -6.3 | -5.4 | -5.7 | -4.8 |
| Coenzyme Q10 | -6.1 | -4.8 | -4.8 | -7.5 | -5.1 | -6.7 | -7.9 | -6.8 | -6.8 | -5.6 | -8.2 | -7.1 | -5.5 |
| Cyanidin | **-9.4** | -7.2 | -8.5 | -8.4 | -7.9 | -7.7 | **-9.6** | -8.9 | -7.9 | -7.6 | -9.2 | -8.2 | -8.8 |
| Daidzein | -8.5 | -7.3 | -7.8 | -8.4 | -7 | -8.6 | -9.3 | -8.2 | -7.7 | -8.3 | -8.3 | -8.7 | -8.2 |
| Delphinidin | -8.1 | -7.9 | -7.6 | -9.0 | -8.6 | -7.2 | -9.6 | -8.3 | **-8.4** | **-9.2** | -9.3 | -8.3 | -8.6 |
| **Ellagic acid** | -7.3 | -7.5 | -8.2 | **-9.4** | -8.4 | **-9.7** | -8.3 | -8.3 | -7.6 | -8.1 | **-10.5** | -8.1 | **-9.7** |
| Epicatechin | -9 | -7.4 | -7.9 | -8.7 | **-9.7** | **-9.5** | -9.5 | -8.3 | -7.5 | -8.9 | -8.9 | -8.2 | -8.2 |
| Epigallocatechin | -7.9 | -7.5 | -7.5 | -8.1 | -6.9 | **-9.6** | -8.3 | -8.3 | -7.5 | -9 | -8.8 | -7.7 | -7.9 |
| **Eriodictyol** | **-9.5** | -7.6 | -7.8 | -8.7 | -7 | **-9.5** | **-9.8** | -8.4 | -7.6 | **-9.4** | **-9.4** | -8.2 | -8.3 |
| Gallic acid | -6.2 | -5.6 | -5.8 | -6 | -6.2 | -6.1 | -6.9 | -6.3 | -5.8 | -5.2 | -6.4 | -5.5 | -6.4 |
| Gallocatechin | -7.9 | -7.5 | -7.4 | -8.2 | -6.9 | -9.6 | -7.4 | -8.2 | -7.4 | -7.6 | -9.2 | -7.8 | -8.4 |
| Genistein | -7.7 | -7.4 | **-8.7** | -8.8 | **-9.6** | -7.2 | -9 | -8.3 | -8 | -7.8 | -7.9 | -9.1 | -8.6 |
| **Glycitein** | -7.4 | **-8.1** | -7.5 | **-9.1** | -6.8 | -7.8 | -7.9 | **-9.4** | -7.2 | -7.3 | -8.1 | **-10.2** | -9.1 |
| Hesperetin | -8.2 | -7.6 | -8.3 | -8.4 | -7.1 | -7.2 | -9.2 | -8.4 | -7.9 | **-9.2** | -9.1 | -8.4 | -8.4 |
| Iodide | -4 | -3.9 | -4.6 | -4.3 | -4.7 | -3.5 | -5.1 | -4.4 | -4.3 | -3.8 | -4.5 | -4.1 | -3.5 |
| Isorhamnetin | **-9.5** | -7 | -7.6 | -8.8 | -7.3 | -8.5 | -8.9 | -7.8 | -8 | -9 | -9.2 | -8.4 | -7.3 |
| Kaempferol | -7.7 | -7.2 | -7.4 | -8.8 | -6.8 | -8.7 | -9.0 | -8.2 | -8.2 | -8.9 | -7.8 | -8.5 | -8.9 |
| Lutein | -8.6 | **-8.4** | -6.5 | -7.5 | -7.8 | -7.7 | -9.3 | -7.6 | -8 | -7.1 | -8.6 | -8.9 | -7.3 |
| **Luteolin** | -8 | -7.6 | **-8.6** | -8.7 | -9.0 | **-9.8** | **-9.7** | -8.5 | -7.8 | **-9.2** | **-9.7** | -8.4 | -8.6 |
| Lycopene | -6.4 | -6.6 | -6.6 | -6.2 | -1.9 | -7.4 | -8.0 | -7.8 | -6.8 | -6.2 | -6.8 | -7.1 | -6.2 |
| Malvidin | -8.5 | -6.8 | -7.3 | -8.5 | -8.2 | -7.5 | -8.7 | -8 | -7.5 | -8.8 | -8.9 | -8.1 | -7.9 |
| Melatonin | -7.5 | -6.5 | -6.9 | -7.1 | -5.5 | -7.9 | -6.1 | -7.2 | -6.4 | -6.8 | -7.9 | -7 | -7.2 |
| Myricetin | -9.2 | -7.1 | -7.6 | -8.9 | -7.1 | -6.5 | -8 | -8.3 | -8 | -8 | -9.4 | -7.9 | -8.5 |
| Naringenin | -8.5 | -7.3 | -7.5 | -8.6 | **-9.5** | **-9.5** | -9 | -8.3 | -7.6 | -9 | -9.1 | -8.2 | -8.5 |
| **Pelargonidin** | -8.3 | -8.1 | -9.1 | **-9.7** | **-10.1** | -6.9 | -8 | **-9.4** | -9 | -7.9 | **-10.1** | -9 | **-10.3** |
| Peonidin | -7.1 | -7.3 | -7.6 | -8.4 | -9.2 | -9 | -9.4 | -8.1 | -7.6 | **-9.2** | -9 | -8.1 | -8.9 |
| Petunidin | -8 | -7.1 | -7.6 | -9.0 | -8.6 | -9.4 | -8.8 | -8.2 | **-8.4** | -7.6 | -9.3 | -8.2 | -8 |
| Pterostilbene | -7.2 | -6.4 | -7 | -7.5 | -6.5 | -8.4 | -7.8 | -7.5 | -6.2 | -7.7 | -7 | -7.6 | -7.5 |
| **Quercetin** | **-9.5** | -7.1 | -7.6 | -8.5 | -7.3 | **-9.4** | -9.2 | -8.5 | **-8.7** | -8.4 | -9.4 | -8.2 | -8.5 |
| Resveratrol | -6.5 | -6.5 | -6.9 | -7.3 | -8.5 | -6.4 | -8.6 | -7.2 | -6.4 | -6.5 | -7.6 | -7.4 | -8 |
| Rosmarinic acid | -7.9 | -7.4 | -5.8 | -8.5 | -7.8 | -7.1 | -8.3 | -8.5 | -5.9 | -8.1 | -9.2 | -7.4 | -8.2 |
| Salicyclic acid | -5.5 | -5.8 | -5.7 | -5.9 | -4.9 | -5.4 | -6.2 | -6.2 | -5.2 | -5.4 | -5.7 | -5.1 | -6.3 |
| Tangeritin | -6.8 | -6.8 | -7.8 | -8 | -6.8 | -9 | -8.2 | -7.8 | -7.6 | -7.4 | -7.1 | -8.2 | -8 |
| **Theaflavin** | **-9.8** | **-8.6** | **-10.8** | **-11.4** | **-8.9** | **-8** | **-11.2** | **-11.3** | **-8.9** | **-9.3** | **-10.2** | **-10.8** | **-11.1** |
| Vitamin C | -5.1 | -5.1 | -5.3 | -5.4 | -5.8 | -4.8 | -6.7 | -6.1 | -5.1 | -5.6 | -5.8 | -5.5 | -5.3 |
| Vitamin E | -6.8 | -5.7 | -7.3 | -5.4 | -4.7 | -5.2 | -4.1 | -5 | -5 | -5.2 | -5.9 | -3.7 | -5 |
| Zeaxanthin | -7.8 | -7.8 | -7.5 | -8 | -8.2 | -6.9 | -7.8 | -8.7 | -8.3 | -7.2 | -8.4 | **-9.6** | -7.9 |
| **Drugs commonly prescribed in Alzheimer disease** | | | | | | | | | | | | | |
| **Donepezil** | **-8.9** | **-6.5** | **-7.5** | **-8.9** | **-9.8** | **-6.7** | **-6.4** | **-8.1** | **-8.6** | **-7.6** | **-7.7** | **-8.6** | **-7.5** |
| Galantamine | -7.4 | -6.5 | -5.7 | -6.8 | -8.0 | -5.8 | -6.4 | -7.7 | -6.7 | -6.8 | -6.8 | -6.3 | -6.5 |
| Rivastigmine | -6.4 | -6.1 | -6.6 | -6.7 | -8.3 | -5.2 | -7.2 | -6.1 | -6.1 | -5.9 | -7.2 | -6.2 | -6.2 |

This method can be defined as filtering out the best using 3 levels of selection, first on the basis of the score (docking), second in addition to ranks the compounds on the basis of interaction with number of receptors and finally eliminating those that violate the minimum criterion for RO5 (Lipinski Rule of Five) and toxicity. The docking interactions were visualized with PYMOL molecular graphics system, version 1.7.4.4 (Schrödinger, LLC, and Portland, OR, USA) and Maestro Visualizer (Maestro, Schrödinger, LLC, New York, NY, 2017.). The docking studies were performed using Autodock4.0[22]. The inhibition constant ($K_i$) of natural compounds against Alzhiemer associated targets was calculated from docking energy using the following equation:

$$Ki = exp(\Delta G * 1000)/RT)$$

Where $\Delta G$ = docking energy; $R = 1.98719$ cal $K^{-1}$ $mol^{-1}$, $T = 298.15°K$, $Ki$ = inhibition constant (nM)

## Results

### Pharmacokinetics Properties

Pharmacokinetics properties of natural compounds to be considered as drug candidates were based on Lipinski's rule of five This rule is formulated for most orally administered drugs, it uses four criteria to determine if a molecule is druglike; to have a molecular weight of $\leq 500$, a LogP (logarithm of partition coefficient) $\leq 5$, five or fewer hydrogen bond donor sites, and ten or fewer hydrogen bond acceptor sites. Molecules violating more than one of these rules may have problems with bioavailability. The entire set of compounds well followed the RO5 (Rule of 5) except 15 of the compounds, out of which eight compounds (*Alpha carotene, Astaxanthin, Beta carotene, Canthaxanthin, Coenzyme Q10, Epigallocatechin, Gallocatechin and Theaflavin*) violating more than one of these rules and seven compounds violationg only single rule *(Chlorogenic acid, Delphinidin, Gallic acid, Lycopene, Myricetin, Vitamin C, Vitamin E )* that created the Lipinski's rule violation by having $molecular\ mass\ > 500$, $logP\ > 5$ and $H - bond\ doner\ > 5$ that can create a problem in oral bioavailability.

TPSA analysis checked the bioavailability of natural compounds, As per the Veber's rule for good oral bioavailability, the number of rotatable bond must be $\leq 10$, and TPSA values $\leq 140\mathring{A}$ [23]. The number of rotatable bonds has been shown to be a very good descriptor of oral bioavailability of drugs and has been found better to discriminate between compounds that have oral bioavailability of drugs. Rotatable bond is defined as any single non-ring bond, bounded to non-terminal heavy (ie, non-hydrogen) atom. Amide C–N bonds are not considered because of their high rotational energy barrier [24]. The numbers of rotatable bonds in all of our compounds were found to be appropriate as in reference compounds (Donepezil, Galantamine and Rivastigmine) except four compounds (*Vit E, Chicoric acid, CoenzymeQ10, Lycopene*) have $> 10\, nRB$. Once TPSA was calculated and it was found that only four compounds (*Chicoric acid, Chlorogenic acid, Rosmarinic acid and Theaflavin* ) have TPSA values ≤ 140Å, then we calculated percentage of absorption for all the 45 compounds using Zhao et al. formula[25].

$$\text{Percentages of Absorption (\%ABS)} = [109 - (0.3345 \times TPSA)]$$

According to above formula, we calculated percentages of absorption of our compounds and as well as our reference compounds. Compounds *(Theaflavin, Chicoric acid, Chlorogenic acid, Myricetin and Rosmarinic acid )* have poor absorbance percentage 36.212%, 37.490%, 52.944%, 59.601% and 59.711% respectively, further details are given in (Table 4).

Table 4: Shows the drug likeness of compounds and violation of Lipinski's rule are highlighted in red colour.

| S. No. | Ligand | Molecular Formula | Molecular Weight (g/mol) | logP | H-Bond Acceptors | H-Bond Donors | TPSA | % ABS | nRB | nAtom | nViolation |
|---|---|---|---|---|---|---|---|---|---|---|---|
| 1 | Alpha carotene | $C_{40}H_{56}$ | 536.44 | 15.187 | 0 | 0 | 0 | 109.00 | 10 | 96 | 2 |
| 2 | Apigenin | $C_{15}H_{10}O_5$ | 270.05 | 1.567 | 1 | 3 | 86.99 | 79.901 | 1 | 30 | 0 |
| 3 | Astaxanthin | $C_{40}H_{52}O_4$ | 596.39 | 9.696 | 4 | 2 | 74.6 | 84.046 | 10 | 96 | 2 |
| 4 | Beta carotene | $C_{40}H_{56}$ | 536.44 | 14.734 | 0 | 0 | 0 | 109.00 | 10 | 96 | 2 |
| 5 | Catechin | $C_{15}H_{14}O_6$ | 290.08 | 0.406 | 1 | 5 | 110.38 | 72.077 | 1 | 35 | 0 |
| 6 | Canthaxanthin | $C_{40}H_{52}O_2$ | 564.4 | 10.788 | 2 | 0 | 34.14 | 97.580 | 10 | 94 | 2 |
| 7 | Chicoric acid | $C_{22}H_{18}O_{12}$ | 472.06 | 0.722 | 8 | 4 | 213.78 | 37.490 | 11 | 50 | 0 |
| 8 | Chlorogenic acid | $C_{16}H_{18}O_9$ | 353.09 | -0.928 | 7 | 5 | 167.58 | 52.944 | 5 | 42 | 1 |
| 9 | Cinnamic acid | $C_9H_8O_2$ | 147.04 | 3.423 | 2 | 0 | 40.13 | 95.576 | 2 | 18 | 0 |
| 10 | Coenzyme Q10 | $C_{59}H_{90}O_4$ | 862.68 | 18.454 | 4 | 0 | 52.6 | 91.405 | 31 | 153 | 2 |
| 11 | Cyanidin | $C_{15}H_{11}O_6^+$ | 287.06 | 0 | 1 | 5 | 101.15 | 75.165 | 1 | 32 | 0 |
| 12 | Daidzein | $C_{15}H_{10}O_4$ | 254.06 | 2.475 | 1 | 2 | 66.76 | 86.668 | 1 | 29 | 0 |
| 13 | Delphinidin | $C_{15}H_{11}O_7^+$ | 303.05 | 0 | 1 | 6 | 121.38 | 68.398 | 1 | 33 | 1 |
| 14 | Ellagic acid | $C_{14}H_6O_8$ | 299.99 | 0.652 | 8 | 2 | 93.06 | 77.871 | 0 | 26 | 0 |
| 15 | Epicatechin | $C_{15}H_{14}O_6$ | 290.08 | 0.406 | 1 | 5 | 110.38 | 72.077 | 1 | 35 | 0 |
| 16 | Epigallocatechin | $C_{15}H_{14}O_7$ | 306.07 | -0.128 | 1 | 6 | 130.61 | 65.310 | 1 | 36 | 2 |
| 17 | Eriodictyol | $C_{15}H_{12}O_6$ | 288.06 | 0.363 | 1 | 4 | 107.22 | 73.134 | 1 | 33 | 0 |
| 18 | Gallic acid | $C_7H_6O_5$ | 169.01 | -0.479 | 2 | 3 | 100.82 | 75.275 | 1 | 17 | 0 |
| 19 | Gallocatechin | $C_{15}H_{14}O_7$ | 306.07 | -0.128 | 1 | 6 | 130.61 | 65.310 | 1 | 36 | 1 |
| 20 | Genistein | $C_{15}H_{10}O_5$ | 270.05 | 1.043 | 5 | 3 | 86.99 | 79.901 | 1 | 30 | 0 |
| 21 | Glycitein | $C_{16}H_{12}O_5$ | 284.07 | 2.031 | 1 | 2 | 75.99 | 83.581 | 2 | 33 | 0 |
| 22 | Hesperetin | $C_{16}H_{14}O_6$ | 302.08 | 0.453 | 1 | 3 | 96.22 | 76.814 | 2 | 36 | 0 |
| 23 | Iodide | I− | 144.14 | 0.035 | 1 | 0 | 17.07 | 103.29 | 4 | 28 | 0 |
| 24 | Isorhamnetin | $C_{16}H_{12}O_7$ | 316.06 | 1.471 | 2 | 4 | 116.45 | 70.047 | 2 | 35 | 0 |
| 25 | Kaempferol | $C_{15}H_{10}O_6$ | 286.05 | 1.915 | 2 | 4 | 107.22 | 73.134 | 1 | 31 | 0 |
| 26 | Lutein | $C_{40}H_{56}O_2$ | 568.43 | 11.283 | 2 | 2 | 40.46 | 95.466 | 10 | 98 | 1 |
| 27 | Luteolin | $C_{15}H_{10}O_6$ | 286.05 | 1.033 | 1 | 4 | 107.22 | 73.134 | 1 | 31 | 0 |
| 28 | Lycopene | $C_{40}H_{56}$ | 536.44 | 14.586 | 0 | 0 | 0 | 109.00 | 16 | 96 | 1 |
| 29 | Malvidin | $C_{17}H_{15}O_7$+ | 331.08 | 0 | 1 | 4 | 99.38 | 75.757 | 3 | 39 | 0 |
| 30 | Melatonin | $C_{13}H_{16}N_2O_2$ | 232.12 | 1.254 | 3 | 2 | 50.36 | 92.154 | 5 | 33 | 0 |
| 31 | Myricetin | $C_{15}H_{10}O_8$ | 318.04 | 0.847 | 2 | 6 | 147.68 | 59.601 | 1 | 33 | 1 |
| 32 | Naringenin | $C_{15}H_{12}O_5$ | 272.07 | 0.897 | 1 | 3 | 86.99 | 79.901 | 1 | 32 | 0 |
| 33 | Pelargonidin | $C_{15}H_{11}O_5^+$ | 271.06 | 0 | 1 | 4 | 80.92 | 81.932 | 1 | 31 | 0 |
| 34 | Peonidin | $C_{16}H_{13}O_6^+$ | 301.07 | 0 | 6 | 4 | 90.15 | 78.844 | 2 | 35 | 0 |
| 35 | Petunidin | $C_{16}H_{13}O_7^+$ ($Cl^-$) | 317.07 | 0 | 1 | 5 | 110.38 | 72.077 | 2 | 36 | 0 |
| 36 | Pterostilbene | $C_{16}H_{16}O_3$ | 256.11 | 2.69 | 3 | 1 | 38.69 | 96.058 | 4 | 35 | 0 |
| 37 | Quercetin | $C_{15}H_{10}O_7$ | 302.04 | 1.381 | 2 | 5 | 127.45 | 66.367 | 1 | 32 | 0 |
| 38 | Resveratrol | $C_{14}H_{12}O_3$ | 228.08 | 3.436 | 0 | 3 | 60.69 | 88.699 | 2 | 29 | 0 |
| 39 | Rosmarinic acid | $C_{18}H_{16}O_8$ | 359.08 | 1.602 | 4 | 4 | 147.35 | 59.711 | 7 | 41 | 0 |
| 40 | Salicyclic acid | $C_7H_6O_3$ | 137.02 | 0.8 | 2 | 1 | 60.36 | 88.809 | 1 | 15 | 0 |
| 41 | Tangeritin | $C_{20}H_{20}O_7$ | 372.12 | 2.236 | 1 | 0 | 72.45 | 84.765 | 6 | 47 | 0 |
| 42 | Theaflavin | $C_{29}H_{24}O_{12}$ | 564.13 | -0.216 | 4 | 9 | 217.6 | 36.212 | 2 | 65 | 2 |
| 43 | Vitamin C | $C_6H_8O_6$ | 175.02 | -1.813 | 6 | 2 | 83.83 | 80.958 | 2 | 19 | 0 |
| 44 | Vitamin E | $C_{29}H_{50}O_2$ | 430.38 | 10.938 | 0 | 1 | 29.46 | 99.145 | 12 | 81 | 1 |
| 45 | Zeaxanthin | $C_{40}H_{56}O_2$ | 568.43 | 10.564 | 2 | 2 | 40.46 | 95.466 | 10 | 98 | 1 |
| **Three Cholinesterase inhibitors are commonly prescribed in Alzheimer disease** | | | | | | | | | | | |
| 1 | Donepezil | $C_{24}H_{30}NO_3^+$ | 380.50 | 3.65 | 1 | 4 | 39.97 | 95.23 | 6 | 58 | 0 |
| 2 | Galantamine | $C_{17}H_{12}NO_3^+$ | 288.36 | 1.54 | 2 | 4 | 43.13 | 94.14 | 1 | 43 | 0 |
| 3 | Rivastigmine | $C_{14}H_{23}N_2O_2^+$ | 251.34 | 2.28 | 1 | 4 | 33.98 | 97.27 | 6 | 41 | 0 |

| | | |
|---|---|---|
| **%ABS-** Percentage of Absorption | **TPSA-** Topological Polar Surface Area | **nrtB-** Number of rotable Atoms |
| **nAtom-** Number of Atoms | **nViolation-** Violation of Lipinski's rule | **XlogP** $\leq 5$ |
| **H-BD** $< 5$ | **H-BA** $< 10$ | **MW** $< 500$ |

## Toxicity Prediction

The computational prediction of toxicities, drug score profiles of natural compounds are promising. An online software PROTOX was used for the prediction of the $LD_{50}$ of the new compounds. Most of the compounds in our study fell in non-toxic zone (above the $1000\ mg/kg$) except Lutein ($10\ mg/kg$), Myricetin ($159\ mg/kg$), Quercetin ($159\ mg/kg$), Salicyclic acid ($480\ mg/kg$) and Zeaxanthin ($10\ mg/kg$), results are shown in the graph in (Fig. 4). The mutagenicity, carcinogenicity and skin irritation properties were also predicted and it wasfound that 05 compounds are mutagenic, 07 compounds are carcinogenic, 38 compounds are very sensitive to skin but 09 of these compounds may cause severe skin irritation problems. Results are listed in (Table 5).

Figure 4 Histogram representation of $LD50$ values of natural compounds compared with commonly prescribes drugs.

Table 5: Shows the toxicity of compounds and various toxicitiness are highlighted in red colour.

| S. no. | Compound | Carcinogenicity | Mutagenicity | Skin Irritation | Skin Sensitization | Biodegradability |
|---|---|---|---|---|---|---|
| 1 | Alpha carotene | Carcinogen | Non mutagen | Severe | Weak | Degradable |
| 2 | Apigenin | Non carcinogen | Non mutagen | None | Strong | Non degradable |
| 3 | Astaxanthin | Non carcinogen | Non mutagen | Mild | Weak | Degradable |
| 4 | Beta carotene | Carcinogen | Non mutagen | Severe | Weak | Degradable |
| 5 | Catechin | Non carcinogen | Non mutagen | None | Strong | Non degradable |
| 6 | Canthaxanthin | Carcinogen | Non mutagen | Severe | Weak | Degradable |
| 7 | Chicoric acid | Non carcinogen | Non mutagen | None | Strong | Degradable |
| 8 | Chlorogenic acid | Non carcinogen | Non mutagen | Mild | None | Degradable |
| 9 | Cinnamic acid | Carcinogen | Non mutagen | Mild | Strong | Degradable |
| 10 | Coenzyme Q10 | Carcinogen | Non mutagen | Severe | None | Degradable |
| 11 | Cyanidin | Non carcinogen | Non mutagen | None | Strong | Non degradable |
| 12 | Daidzein | Non carcinogen | Non mutagen | None | Strong | Non degradable |
| 13 | Delphinidin | Non carcinogen | Non mutagen | None | Strong | Non degradable |
| 14 | Ellagic acid | Non carcinogen | Non mutagen | Mild | Weak | Non degradable |
| 15 | Epicatechin | Non carcinogen | Non mutagen | None | Strong | Non degradable |
| 16 | Epigallocatechin | Non carcinogen | Non mutagen | None | Strong | Non degradable |
| 17 | Eriodictyol | Non carcinogen | Non mutagen | None | Strong | Non degradable |
| 18 | Gallic acid | Non carcinogen | Non mutagen | None | Strong | Degradable |
| 19 | Gallocatechin | Non carcinogen | Non mutagen | None | Strong | Non degradable |
| 20 | Genistein | Non carcinogen | Non mutagen | None | Strong | Non degradable |
| 21 | Glycitein | Non carcinogen | Non mutagen | None | None | Degradable |
| 22 | Hesperetin | Non carcinogen | Non mutagen | None | Strong | Degradable |
| 23 | Iodide | Non carcinogen | Non mutagen | Severe | Weak | Degradable |
| 24 | Isorhamnetin | Non carcinogen | Mutagen | None | Strong | Non degradable |
| 25 | Kaempferol | Non carcinogen | Mutagen | None | Strong | Non degradable |
| 26 | Lutein | Non carcinogen | Non mutagen | Severe | Weak | Degradable |
| 27 | Luteolin | Non carcinogen | Non mutagen | None | Strong | Non degradable |
| 28 | Lycopene | Carcinogen | Non mutagen | Severe | Weak | Degradable |
| 29 | Malvidin | Non carcinogen | Non mutagen | None | Strong | Non degradable |
| 30 | Melatonin | Non carcinogen | Non mutagen | None | None | Non degradable |
| 31 | Myricetin | Non carcinogen | Mutagen | None | Strong | Non degradable |
| 32 | Naringenin | Non carcinogen | Non mutagen | None | Strong | Non degradable |
| 33 | Pelargonidin | Non carcinogen | Non mutagen | None | Strong | Non degradable |
| 34 | Peonidin | Non carcinogen | Non mutagen | None | Strong | Non degradable |
| 35 | Petunidin | Non carcinogen | Non mutagen | None | Strong | Non degradable |
| 36 | Pterostilbene | Non carcinogen | Mutagen | None | Strong | Non degradable |
| 37 | Quercetin | Non carcinogen | Non mutagen | None | Strong | Non degradable |
| 38 | Resveratrol | Non carcinogen | Non mutagen | None | Strong | Non degradable |
| 39 | Rosmarinic acid | Non carcinogen | Non mutagen | None | Strong | Non degradable |
| 40 | Salicyclic acid | Non carcinogen | Non mutagen | None | None | Degradable |
| 41 | Tangeritin | Non carcinogen | Non mutagen | None | Strong | Degradable |
| 42 | Theaflavin | Non carcinogen | Non mutagen | None | Strong | Non degradable |
| 43 | Vitamin C | Non carcinogen | Non mutagen | None | None | Degradable |
| 44 | Vitamin E | Carcinogen | Non mutagen | Severe | None | Non degradable |
| 45 | Zeaxanthin | Non carcinogen | Non mutagen | Severe | Weak | Degradable |

## Molecular docking Studies

To ensure the interaction between the natural compounds and Alzheimer disease associated targets, we performed molecular docking analysis using Autodock4.2. Each of the compounds was docked with all the 13 Alzheimer disease associated targets individually. These compounds showed very good binding affinity with all the 13 Alzheimer associated targets.

Among all of the compounds only 09 compounds interacted with maximum number of targets, such as $\boldsymbol{Theaflavin}$ showed maximum interactions, as it interacted with 11 Alzheimer associated targets ($1EQG$, $1MX1$, $1PBQ$, $1Q5K$, $2FV5$, $3BKL$, $3G9N$, $3QMO$, $4B0P$, $4DJU$ and $4EY5$); whereas compound $\boldsymbol{Catechin}$ interacted with 10 targets ($1MX1$, $1PBQ$, $1Q5K$, $1QWC$, $2FV5$, $3BKL$, $3G9N$, $4B0P$, $4DJU$ *and* $4EY5$), $\boldsymbol{Pelargonidin}$ interacted with 05 targets ($1Q5K$, $1QWC$, $3BKL$, $4B0P$ and $4EY5$), $\boldsymbol{Luteolin}$ intercalated with 05 targets ($1PBQ$, $1UDT$, $2FV5$, $3QMO$ and $4B0P$),

***Ellagic acid*** compound interacted with 04 targets ($1Q5K$, $1UDT$, $4B0P$ and $4EY5$), ***Eriodictyol*** intercalated with 04 targets ($1EQG$, $1UDT$, $2FV5$ and $3QMO$), ***Glycitein*** intercalated with 04 targets ($1MX1$, $1Q5K$, $3BKL$ and $4DJU$), ***Quercetin*** intercalated with 03 targets ($1EQG$, $1UDT$ and $3G9N$) and ***Astaxanthin*** intercalated with 03 targets ($3BKL$, $4DJU$ and $4EY5$). Even though compounds ***Astaxanthin*** and ***Theaflavin*** showed very good binding affinity, we did not consider them for further docking analysis because they did not comply with the Lipinski's rule. ***Astaxanthin*** has greater molecular mass (596.39 Daltons) and greater log P (9.696) value, whereas ***Theaflavin*** has greater molecular mass (564.13 g/mol) and more than five H-bond donor. The range of the binding affinities of all the compounds lies between $-4.00\, kcal/mol$ to $-11.4\, kcal/mol$. In addition, the residues surrounding the ligands (within 4 Å) in all docking results provide short range polar interactions that stabilize the formation of complex (with the help of delocalization of charges on ligands). It is not possible to show all the binding poses but we have shown 07 poses which have maximum binding affinity amoung all of them (Fig. 5). The inhibitor constant ($K_i$) is an indicator of the effectiveness of an inhibitor, with a greatly potent inhibitor being indicated by a low $K_i$ because smaller the $K_i$, the greater the binding affinity and the smaller amount of medication needed in order to inhibit the activity of that enzyme [34]. All of our compounds were found to have a very small $Ki$ value except compound *Cinnamic acid*, (with *1EQG, 1PBQ, 1QWC, 2FV5, 3G9N, 4B0P and 4EY5*), *Chlorogenic acid* (*with* $1QWC$), *Coenzyme* $Q10$ (with *1MX1, 1PBQ* and *1QWC), Gallic acid* (with *3QMO*), Iodide (with all targets), *Lycopene* (with *1QWC*), *Salicyclic acid* (with *1QWC, 1UDT, 3G9N, 3QMO and 4DJU*), *Vitamin C* (with *1EQG, 1MX1, 1PBQ, 1Q5K, 1UDT, 3G9N* and *4EY5*), *Vitamin E* (with *1Q5K, 1QWC, 1UDT, 2FV5, 3BKL, 3G9N, 3QMO, 4DJU* and *4EY5*) and all three drugs commonly prescribed in Alzheimer disease (*Donepezil, Galantamine* and *Rivastigmine*) that have very poor $K_i$ values with *4DJU* (Table 6). In this way, all of these results suggest that our natural compounds potentially interfere with Alzheimer associated targets, which should prompt further investigations to expose the mechanism of our compounds against Alzheimer disease in vivo.

**Figure 5** Molecular docking poses of the top seven representative compounds: Crystallographic structure of **(A)** ***Catechin*** complexed to *Butyrylcholinesterase(BuChE)(*Pdb id: 4BOP), **(B)** ***Ellagic*** acid complexed to *Butyrylcholinesterase (BuChE)* (Pdb id: 4BOP), **(C)** ***Eriodictyol*** complexed to *TNF-α converting enzyme (TACE)*(Pdb id: 2FV5), **(D)** ***Glycitein*** complexed to *β-Site amyloid precursor protein* (Pdb id: 4DJU), **(E)** ***Luteolin*** complexed to *TNF-α converting enzyme* (Pdb id: 2FV5), **(F)** ***Pelagonidin*** complexed to *Acetylcholinesterase (AChE)* (Pdb id: 4EY5), **(G)** ***Quercetin*** complexed to *Cyclooxygenase-1 (COX-1)* (Pdb id: 1EQG).

**Table 6:** Calculated $Ki(nM)$ value from docking energy of 45 natural compounds against 13 Alzheimer disease associated targets. Compounds that have very poor $Ki$ values are highlighted in grey colour.

| **Compound Names** | **1EQG** | **1MX1** | **1PBQ** | **1Q5K** | **1QWC** | **1UDT** | **2FV5** | **3BKL** | **3G9N** | **3QMO** | **4B0P** | **4DJU** | **4EY5** |
|---|---|---|---|---|---|---|---|---|---|---|---|---|---|
| Alpha carotene | 4.46E-06 | 4.46E-06 | 5.28E-06 | 4.97E-07 | 4.46E-06 | 3.18E-06 | 1.04E-05 | 2.69E-06 | 3.18E-06 | 3.76E-06 | 1.37E-06 | 3.54E-07 | 6.25E-06 |
| Apigenin | 2.14E-07 | 1.52E-07 | 6.96E-07 | 4.20E-07 | 1.52E-07 | 1.29E-07 | 1.37E-06 | 8.24E-07 | 1.92E-06 | 3.76E-06 | 1.62E-06 | 6.96E-07 | 2.99E-07 |
| Astaxanthin | 9.76E-07 | 1.62E-06 | 3.76E-06 | 4.20E-07 | 2.85E-05 | 7.39E-06 | 1.37E-06 | 4.68E-08 | 1.16E-06 | 3.76E-06 | 4.20E-07 | 1.29E-07 | 1.80E-07 |
| Beta carotene | 1.62E-06 | 6.25E-06 | 4.20E-07 | 3.76E-06 | 2.53E-07 | 2.27E-06 | 2.69E-06 | 1.92E-06 | 4.46E-06 | 1.62E-06 | 5.88E-07 | 1.52E-07 | 6.25E-06 |
| Canthaxanthin | 6.96E-07 | 1.62E-06 | 1.92E-06 | 1.16E-06 | 1.80E-07 | 5.28E-06 | 9.76E-07 | 2.69E-06 | 4.20E-07 | 3.18E-06 | 1.62E-06 | 3.76E-06 | 8.75E-06 |
| **Catechin** | 4.20E-07 | 1.29E-07 | 5.88E-07 | 7.76E-08 | 5.54E-08 | 2.27E-06 | 3.95E-08 | 2.38E-08 | 2.53E-07 | 2.69E-06 | **1.70E-08** | 1.09E-07 | 9.19E-08 |
| Chicoric acid | 1.37E-06 | 1.45E-05 | 4.46E-06 | 5.88E-07 | 4.97E-07 | 6.25E-06 | 5.88E-07 | 4.97E-07 | 3.76E-06 | 5.28E-06 | 9.19E-08 | 4.46E-06 | 9.76E-07 |
| Chlorogenic acid | 2.69E-06 | 3.76E-06 | 8.75E-06 | 5.88E-07 | 0.000303 | 1.72E-05 | 6.55E-08 | 1.92E-06 | 3.18E-06 | 3.18E-06 | 8.24E-07 | 3.76E-06 | 2.69E-06 |
| Cinnamic acid | 0.00021625 | 9.30E-05 | 0.000834 | 2.85E-05 | 0.000359 | 1.72E-05 | 0.000359 | 3.38E-05 | 0.000359 | 2.41E-05 | 0.00011 | 6.64E-05 | 0.000303 |
| Coenzyme Q10 | 3.38E-05 | 0.000303 | 0.000303 | 3.18E-06 | 0.000183 | 1.23E-05 | 1.62E-06 | 1.04E-05 | 1.04E-05 | 7.85E-05 | 9.76E-07 | 6.25E-06 | 9.30E-05 |
| Cyanidin | 1.29E-07 | 5.28E-06 | 5.88E-07 | 6.96E-07 | 1.62E-06 | 2.27E-06 | 9.19E-08 | 2.99E-07 | 1.62E-06 | 2.69E-06 | 1.80E-07 | 9.76E-07 | 3.54E-07 |
| Daidzein | 5.88E-07 | 4.46E-06 | 1.92E-06 | 6.96E-07 | 7.39E-06 | 4.97E-07 | 1.52E-07 | 9.76E-07 | 2.27E-06 | 8.24E-07 | 8.24E-07 | 4.20E-07 | 9.76E-07 |
| Delphinidin | 1.16E-06 | 1.62E-06 | 2.69E-06 | 2.53E-07 | 4.97E-07 | 5.28E-06 | 9.19E-08 | 8.24E-07 | 6.96E-07 | 1.80E-07 | 1.52E-07 | 8.24E-07 | 4.97E-07 |
| **Ellagic acid** | 4.46E-06 | 3.18E-06 | 9.76E-07 | 1.29E-07 | 6.96E-07 | 7.76E-08 | 8.24E-07 | 8.24E-07 | 2.69E-06 | 1.16E-06 | **2.01E-08** | 1.16E-06 | 7.76E-08 |
| Epicatechin | 2.53E-07 | 3.76E-06 | 1.62E-06 | 4.20E-07 | 7.76E-08 | 1.09E-07 | 1.09E-07 | 8.24E-07 | 3.18E-06 | 2.99E-07 | 2.99E-07 | 9.76E-07 | 9.76E-07 |
| Epigallocatechin | 1.62E-06 | 3.18E-06 | 3.18E-06 | 1.16E-06 | 8.75E-06 | 9.19E-08 | 8.24E-07 | 8.24E-07 | 3.18E-06 | 2.53E-07 | 3.54E-07 | 2.27E-06 | 1.62E-06 |
| **Eriodictyol** | 1.09E-07 | 2.69E-06 | 1.92E-06 | 4.20E-07 | 7.39E-06 | 1.09E-07 | **6.55E-08** | 6.96E-07 | 2.69E-06 | 1.29E-07 | 1.29E-07 | 9.76E-07 | 8.24E-07 |
| Gallic acid | 2.85E-05 | 7.85E-05 | 5.60E-05 | 4.00E-05 | 2.85E-05 | 3.38E-05 | 8.75E-06 | 2.41E-05 | 5.60E-05 | 0.000154 | 2.04E-05 | 9.30E-05 | 2.04E-05 |
| Gallocatechin | 1.62E-06 | 3.18E-06 | 3.76E-06 | 9.76E-07 | 8.75E-06 | 9.19E-08 | 3.76E-06 | 9.76E-07 | 3.76E-06 | 2.69E-06 | 1.80E-07 | 1.92E-06 | 6.96E-07 |
| Genistein | 2.27E-06 | 3.76E-06 | 4.20E-07 | 3.54E-07 | 9.19E-08 | 5.28E-06 | 2.53E-07 | 8.24E-07 | 1.37E-06 | 1.92E-06 | 1.62E-06 | 2.14E-07 | 4.97E-07 |
| Glycitein | 3.76E-06 | 1.16E-06 | 3.18E-06 | 2.14E-07 | 1.04E-05 | 1.92E-06 | 1.62E-06 | 1.29E-07 | 5.28E-06 | 4.46E-06 | 1.16E-06 | **3.34E-08** | 2.14E-07 |
| Hesperetin | 9.76E-07 | 2.69E-06 | 8.24E-07 | 6.96E-07 | 6.25E-06 | 5.28E-06 | 1.80E-07 | 6.96E-07 | 1.62E-06 | 1.80E-07 | 2.14E-07 | 6.96E-07 | 6.96E-07 |
| Iodide | 0.00116939 | 0.001384 | 0.000425 | 0.000705 | 0.000359 | 0.002719 | 0.000183 | 0.000595 | 0.000705 | 0.001639 | 0.000503 | 0.000988 | 0.002719 |
| Isorhamnetin | 1.09E-07 | 7.39E-06 | 2.69E-06 | 3.54E-07 | 4.46E-06 | 5.88E-07 | 2.99E-07 | 1.92E-06 | 1.37E-06 | 2.53E-07 | 1.80E-07 | 6.96E-07 | 4.46E-06 |
| Kaempferol | 2.27E-06 | 5.28E-06 | 3.76E-06 | 3.54E-07 | 1.04E-05 | 4.20E-07 | 2.53E-07 | 9.76E-07 | 9.76E-07 | 2.99E-07 | 1.92E-06 | 5.88E-07 | 2.99E-07 |
| Lutein | 4.97E-07 | 6.96E-07 | 1.72E-05 | 3.18E-06 | 1.92E-06 | 2.27E-06 | 1.52E-07 | 2.69E-06 | 1.37E-06 | 6.25E-06 | 4.97E-07 | 2.99E-07 | 4.46E-06 |
| Luteolin | 1.37E-06 | 2.69E-06 | 4.97E-07 | 4.20E-07 | 2.53E-07 | **6.55E-08** | 7.76E-08 | 5.88E-07 | 1.92E-06 | 1.80E-07 | 7.76E-08 | 6.96E-07 | 4.97E-07 |
| Lycopene | 2.04E-05 | 1.45E-05 | 1.45E-05 | 2.85E-05 | 0.040484 | 3.76E-06 | 1.37E-06 | 1.92E-06 | 1.04E-05 | 2.85E-05 | 1.04E-05 | 6.25E-06 | 2.85E-05 |
| Malvidin | 5.88E-07 | 1.04E-05 | 4.46E-06 | 5.88E-07 | 9.76E-07 | 3.18E-06 | 4.20E-07 | 1.37E-06 | 3.18E-06 | 3.54E-07 | 2.99E-07 | 1.16E-06 | 1.62E-06 |
| Melatonin | 3.18E-06 | 1.72E-05 | 8.75E-06 | 6.25E-06 | 9.30E-05 | 1.62E-06 | 3.38E-05 | 5.28E-06 | 2.04E-05 | 1.04E-05 | 1.62E-06 | 7.39E-06 | 5.28E-06 |
| Myricetin | 1.80E-07 | 6.25E-06 | 2.69E-06 | 2.99E-07 | 6.25E-06 | 1.72E-05 | 1.37E-06 | 8.24E-07 | 1.37E-06 | 1.37E-06 | 1.29E-07 | 1.62E-06 | 5.88E-07 |
| Naringenin | 5.88E-07 | 4.46E-06 | 3.18E-06 | 4.97E-07 | 1.09E-07 | 1.09E-07 | 2.53E-07 | 8.24E-07 | 2.69E-06 | 2.53E-07 | 2.14E-07 | 9.76E-07 | 5.88E-07 |
| Pelargonidin | 8.24E-07 | 1.16E-06 | 2.14E-07 | 7.76E-08 | 3.95E-08 | 8.75E-06 | 1.37E-06 | 1.29E-07 | 2.53E-07 | 1.62E-06 | 3.95E-08 | 2.53E-07 | **2.82E-08** |
| Peonidin | 6.25E-06 | 4.46E-06 | 2.69E-06 | 6.96E-07 | 1.80E-07 | 2.53E-07 | 1.29E-07 | 1.16E-06 | 2.69E-06 | 1.80E-07 | 2.53E-07 | 1.16E-06 | 2.99E-07 |
| Petunidin | 1.37E-06 | 6.25E-06 | 2.69E-06 | 2.53E-07 | 4.97E-07 | 1.29E-07 | 3.54E-07 | 9.76E-07 | 6.96E-07 | 2.69E-06 | 1.52E-07 | 9.76E-07 | 1.37E-06 |
| Pterostilbene | 5.28E-06 | 2.04E-05 | 7.39E-06 | 3.18E-06 | 1.72E-05 | 6.96E-07 | 1.92E-06 | 3.18E-06 | 2.85E-05 | 2.27E-06 | 7.39E-06 | 2.69E-06 | 3.18E-06 |
| Quercetin | **1.09E-07** | 6.25E-06 | 2.69E-06 | 5.88E-07 | 4.46E-06 | **1.09E-07** | 1.80E-07 | 5.88E-07 | 4.20E-07 | 6.96E-07 | 1.29E-07 | 9.76E-07 | 5.88E-07 |
| Resveratrol | 1.72E-05 | 1.72E-05 | 8.75E-06 | 4.46E-06 | 5.88E-07 | 2.04E-05 | 4.97E-07 | 5.28E-06 | 2.04E-05 | 1.72E-05 | 2.69E-06 | 3.76E-06 | 1.37E-06 |
| Rosmarinic acid | 1.62E-06 | 3.76E-06 | 5.60E-05 | 5.88E-07 | 1.92E-06 | 6.25E-06 | 8.24E-07 | 5.88E-07 | 4.73E-05 | 1.16E-06 | 1.80E-07 | 3.76E-06 | 9.76E-07 |
| Salicyclic acid | 9.30E-05 | 5.60E-05 | 6.64E-05 | 4.73E-05 | 0.000256 | 0.00011 | 2.85E-05 | 2.85E-05 | 0.000154 | 0.00011 | 6.64E-05 | 0.000183 | 2.41E-05 |
| Tangeritin | 1.04E-05 | 1.04E-05 | 1.92E-06 | 1.37E-06 | 1.04E-05 | 2.53E-07 | 9.76E-07 | 1.92E-06 | 2.69E-06 | 3.76E-06 | 6.25E-06 | 9.76E-07 | 1.37E-06 |
| Theaflavin | 6.55E-08 | 4.97E-07 | 1.21E-08 | 4.40E-09 | 2.99E-07 | 1.37E-06 | 6.17E-09 | 5.21E-09 | 2.99E-07 | 1.52E-07 | 3.34E-08 | 1.21E-08 | 7.30E-09 |
| Vitamin C | 0.00018266 | 0.000183 | 0.00013 | 0.00011 | 5.60E-05 | 0.000303 | 1.23E-05 | 3.38E-05 | 0.000183 | 7.85E-05 | 5.60E-05 | 9.30E-05 | 0.00013 |
| Vitamin E | 1.04E-05 | 6.64E-05 | 4.46E-06 | 0.00011 | 0.000359 | 0.000154 | 0.000988 | 0.000216 | 0.000216 | 0.000154 | 4.73E-05 | 0.00194 | 0.000216 |
| Zeaxanthin | 1.92E-06 | 1.92E-06 | 3.18E-06 | 1.37E-06 | 9.76E-07 | 8.75E-06 | 1.92E-06 | 4.20E-07 | 8.24E-07 | 5.28E-06 | 6.96E-07 | 9.19E-08 | 1.62E-06 |
| **Drugs commonly prescribed in Alzheimer disease** | | | | | | | | | | | | | |
| Donepezil | 3.18E-06 | 4.46E-06 | 5.28E-06 | 4.97E-07 | 4.46E-06 | 3.18E-06 | 1.04E-05 | 2.69E-06 | 3.18E-06 | 3.76E-06 | 1.37E-06 | 1.000001 | 6.25E-06 |
| Galantamine | 1.29E-07 | 1.52E-07 | 6.96E-07 | 4.20E-07 | 1.52E-07 | 1.29E-07 | 1.37E-06 | 8.24E-07 | 1.92E-06 | 3.76E-06 | 1.62E-06 | 1.000001 | 2.99E-07 |
| Rivastigmine | 7.39E-06 | 1.62E-06 | 3.76E-06 | 4.20E-07 | 2.85E-05 | 7.39E-06 | 1.37E-06 | 4.68E-08 | 1.16E-06 | 3.76E-06 | 4.20E-07 | 1.000000 | 1.80E-07 |

## Discussion

Natural products have been used since ancient times and are well recognized as sources of drugs in several human ailments. The healing ability of these herbs and medicinal plants draw attention to study natural products as a potentially valuable resource of drug molecules, they are evolutionarily optimized as drug-like molecules and remain the best sources of drugs and drug leads [39][40]. In our study, we chose 45 natural compounds that have remarkable antioxidant property and act mainly by scavenging free radical species. In recent years, significant data have been gathered, indicating the level of oxidative stress increase in the brain in AD condition. This may have a role in the pathogenesis of neuron degeneration and death [41]. Thus, protection and inhibition against oxidative stress may be considered an important critarion in the development of anti-Alzheimer agents. Therefore, treatment with these antioxidant agents might prevent or reduce the progression of AD. These natural compounds may prove to be novel anti-Alzheimer agents. In order to exploit all the properties for a compound to behave as a drug, the study was done in silico by using different computational tools based on Chemo-informatics or Bioinformatics. The drug-like property predictions showed that most of the compounds followed the Lipinski's rule of five and ADMET. Finally, it can be seen from our docking studies also that the compounds exhibit a strong interaction with Alzheimer associated targets. As this is an insilico study, therefore one should not forget to consider the complex biological metabolic processes during the onset of Alzheimer. The interaction parameter may vary due to multiple interactions in complex system which is another subject of consideration. These results suggest that these compounds may be considered to be a kind of novel anti-Alzheimer's Disease drug agents, which are multi target-directed ligands with not only antioxidant activity but also inhibitory and neuroprotective activities. Molecular docking and pharmacokinetics studies showed that most of our compounds fulfil the requirements for an anti-Alzheimer drug, such as ADMET, RO5, non toxicity, binding affinity, inhibition constants, antioxidant, and neuroprotective inhibitory properties and good interaction with Alzheimer associate targets. In this way these results suggest that natural compounds potentially interfere with Alzheimer associated targets, which should prompt further investigations to expose the mechanism of our compounds against Alzheimer disease in vivo. Therefore, we recommend them for future in vivo studies and possible clinical trials.

# Conclusion

Several natural products are used alone or in combination with other neuroprotective compounds to improve memory and cognition in AD patients as supported by various experimental studies. Altogether, this pioneering study was used to preliminarily investigate the potential compounds (drug candidates) from natural products and conventional docking study to analyze the best docked ligands permitted us to know the binding mode of compounds. Binding energies of the drug–targets interactions are important to describe how fit the drug binds to the target macromolecule (13 Alzheimer disease associated targets). Further studies are essential to explore the target specific effect of these natural compounds on various signalling pathways, mode of action in various brain regions, the ability to cross the blood brain barrier and the mechanism behind the synergistic action of the antioxidant agents on the target. Using novel Pharmaceutical engineering and medicinal chemistry approach to prepare novel formulations or design new compounds based on natural templates, opens up a new window into using natural therapeutic agents against AD.

**Acknowledgements**

The authors are grateful to the Centre for Interdisciplinary Research in Basic Sciences (CIRBSc), Jamia Millia Islamia for providing the research infrastructure.

Aftab Alam is supported by a research fellowship of the University Grants Commission, Government of India., SA and MZM are supported by the Indian Council of Medical Research under SRF (Senior Research Fellowship).

**Disclosure**

The authors report no conflicts of interests in this work.

Figure-1

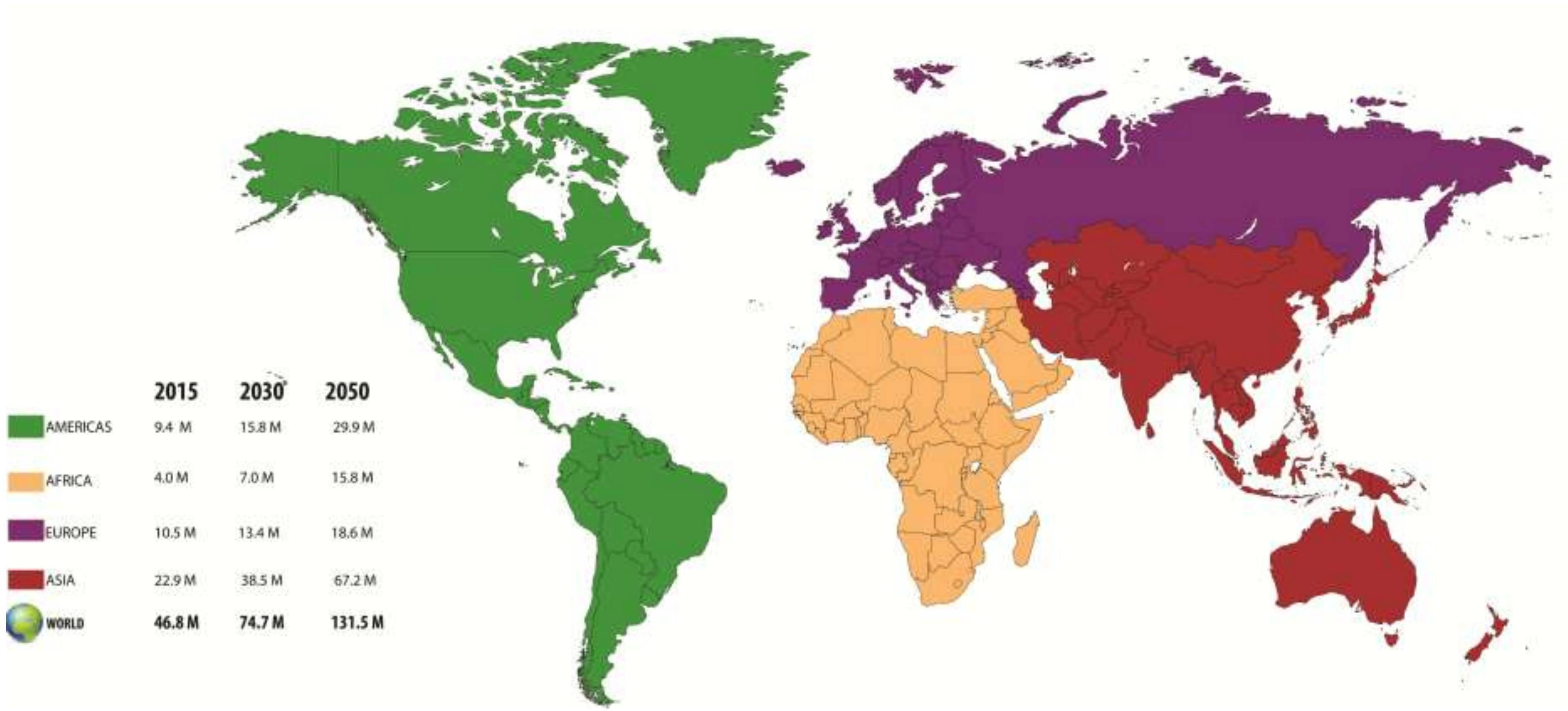

Figure-2

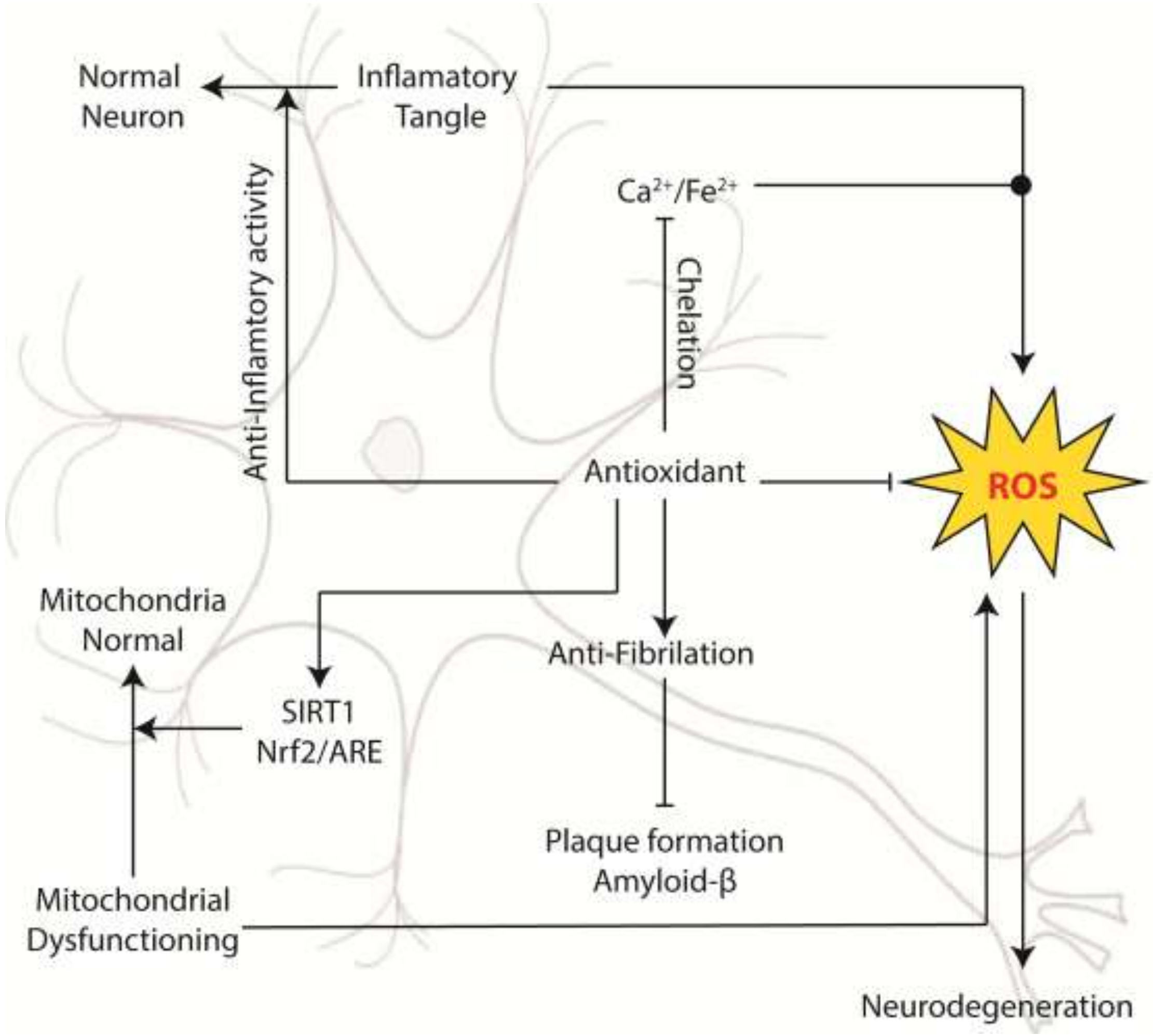

Figure-3

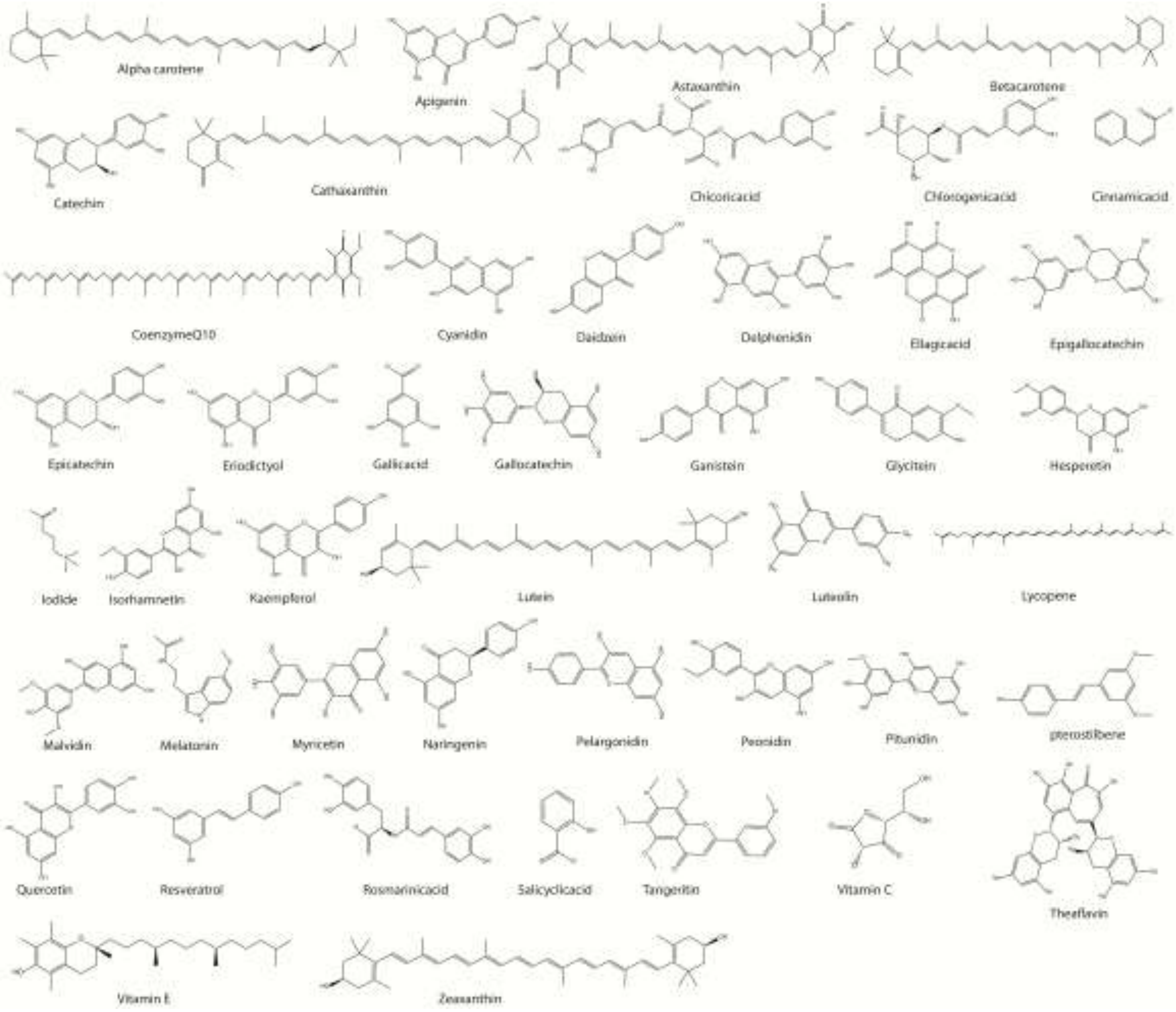

Figure-4

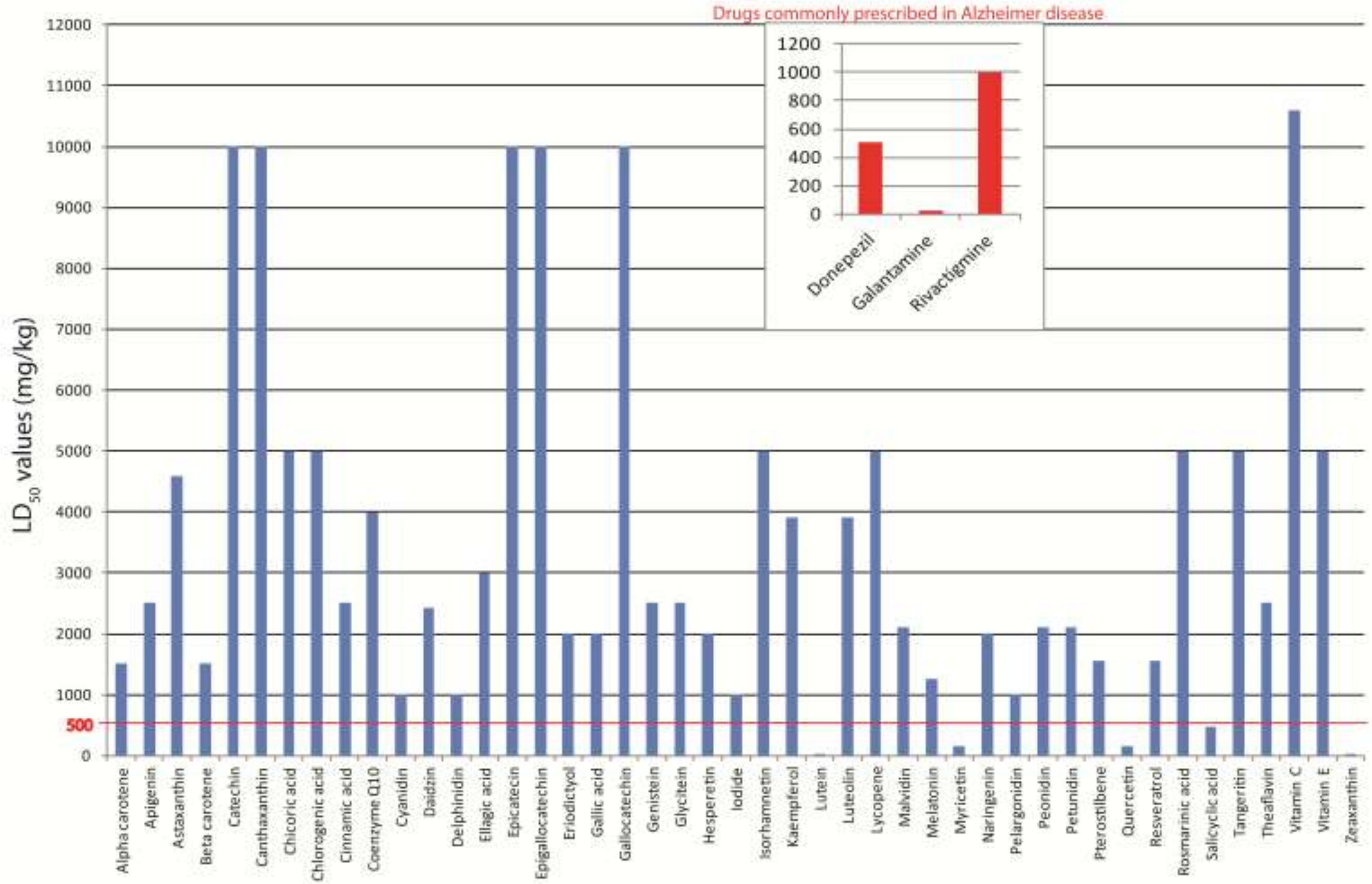

Figure-5

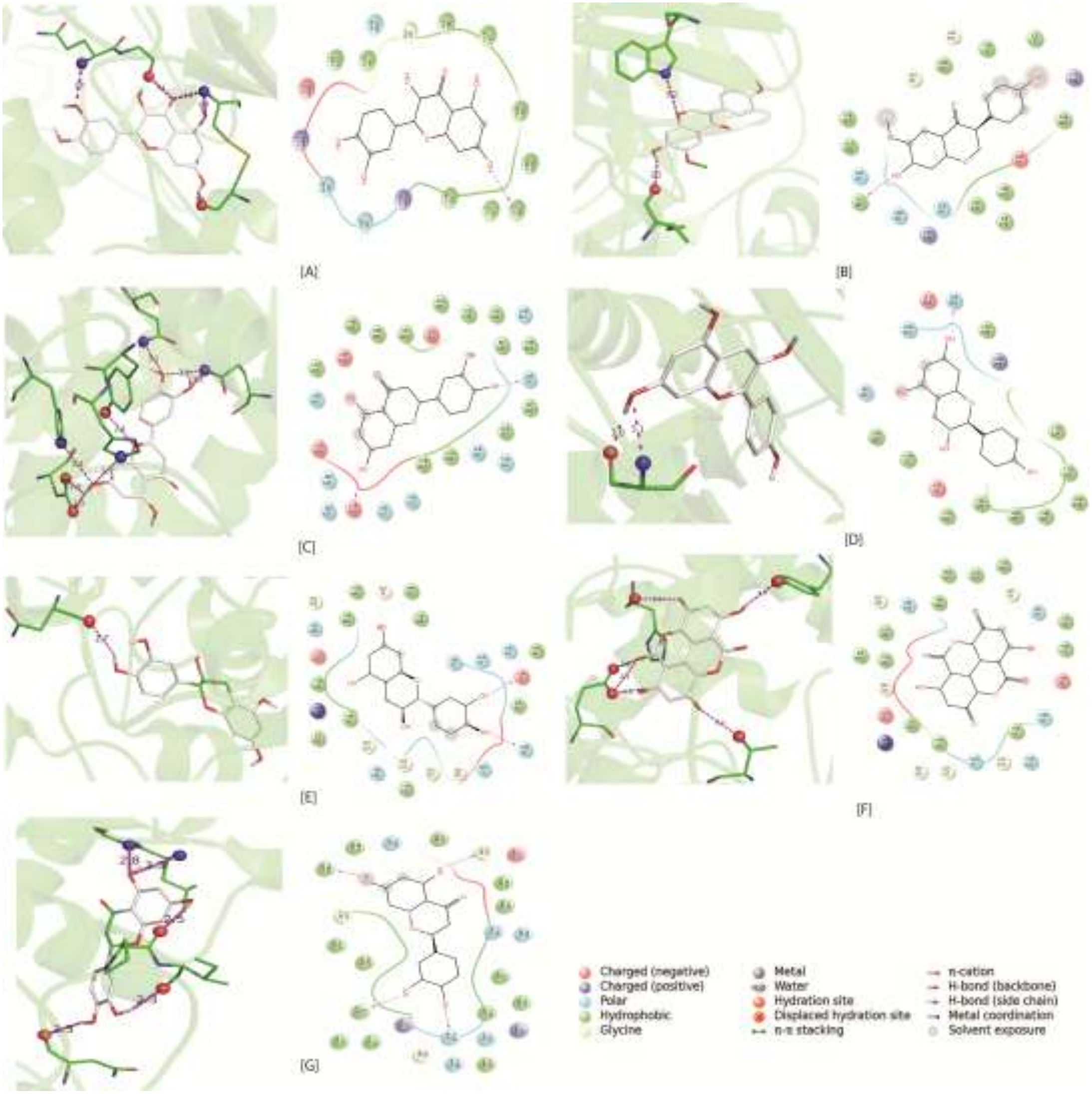